\documentclass[11pt]{article}

\usepackage{geometry}                
\usepackage{graphicx}
\newcommand{\be}{\begin{equation}}
\newcommand{\ee}{\end{equation}}
\newcommand{\bea}{\begin{eqnarray}}
\newcommand{\eea}{\end{eqnarray}}
\newcommand{\beas}{\begin{eqnarray*}}
\newcommand{\eeas}{\end{eqnarray*}}

\usepackage{amssymb}
\usepackage{amsmath}
\usepackage{amsfonts}
\usepackage{graphicx}
\usepackage{epstopdf}

\begin{document}

\title{Heterogeneity, correlations and financial contagion}
\author{Fabio Caccioli$^{1}$, Thomas A. Catanach$^{2}$, and J. Doyne Farmer$^{1}$
~~~~~\\
{\em 1 - Santa Fe Institute, 1399 Hyde Park road, Santa Fe, NM 87501, USA }\\
{\em 2 - University of Notre Dame, Notre Dame, IN 46556 USA}\\ 
}

\maketitle
\begin{abstract}
We consider a model of contagion in financial networks recently introduced in \cite{GK10}, and we characterize the effect of a few features empirically observed in real networks on the stability of the system. Notably, we consider the effect of heterogeneous degree distributions, heterogeneous balance sheet size and degree correlations between banks. We study the probability of contagion conditional on the failure of a random bank, the most connected bank and the biggest bank, and we consider the effect of targeted policies aimed at increasing the capital requirements of a few banks with high connectivity or big balance sheets. Networks with heterogeneous degree distributions are shown to be more resilient to contagion triggered by the failure of a random bank, but more fragile with respect to contagion triggered by the failure of highly connected nodes. A power law distribution of balance sheet size is shown to induce an inefficient diversification that makes the system more prone to contagion events. A targeted policy aimed at reinforcing the stability of the biggest banks is shown to improve the stability of the system in the regime of high average degree.  Finally, disassortative mixing, such as that observed in real banking networks, is shown to enhance the stability of the system.

\end{abstract}

\section{Introduction}
The recent economic crisis,  which began with the burst of the housing market bubble in the U.S. and then propagated across different financial sectors all over the world \cite{TCJ08},  has highlighted the networked structure of the economic world. 
Contemporary finance is clearly  characterized by a high level of connectivity between financial institutions \cite{emp_italy,emp_italy2,emp_Austria,emp_Brasil10}.  In stable market conditions such connections allow financial institutions (in this case banks) to diversify their investment and to safely account for liquidity requests from their investors without facing the risk of insolvency due to temporary shortage of liquid assets.
The same networked structure may, however, become a channel of contagion and stress amplification when some institutions go bankrupt. In this case, banks linked to those that fail may be forced to revise the value of their assets and may face the risk of going bankrupt too.\\

One of the key problems in this context is that of understanding the role of the network structure in relation to the contagion effect, and a growing part of the literature in the field has been recently devoted to address this issue \cite{GK10, NYYA08,CM09,HM11, ACM10,AG01,Georg10}.
In particular, we focus here on the attempt to model the propagation of failures in a financial system as an epidemic spreading process in a network of interlinked balance-sheets \cite{GK10,NYYA08,HM11}. With this approach the assets of a bank correspond to the liabilities of other banks, and troubles may arise  if a bank goes bankrupt and is not able to pay back the loans obtained from other banks. These are then forced to revise their balance sheets and may go bankrupt if the amount of their assets becomes smaller than their liabilities. Starting from a small number of failed banks, the aim is to characterize the probability that failures propagate at the systemic level as a function of some relevant parameters, like the connectivity of the network and the capital buffer of banks.\\

Here we consider the model introduced by Gai and Kapadia in \cite{GK10}.  They consider a simple (yet rich) model of interlinked balance-sheets and study it with tools borrowed from the theory of complex networks\footnote{The model considered in \cite{GK10} is actually a generalization of the model for cascades in random networks introduced in \cite{Watts02}.}. In the present paper we extend the analysis made in \cite{GK10} to account for properties empirically observed in real financial networks, notably heterogeneous degree distributions, heterogeneous assets distributions and degree correlations among connected banks. 
Our major findings are:
\begin{itemize}
\item Heterogeneous networks are more robust with respect to the failure of random banks but more fragile with respect to the failure of the most connected banks.
\item A heterogeneous distribution of assets increases the probability of contagion even with respect to random failures.
\item Within the framework of the model a policy targeted to increase the capital buffer of a few big banks can significantly reduce the contagion probability in highly connected networks.  In contrast, a policy targeted to increase the capital buffer of the most connected banks is ineffective.
\item Correlations, depending on their direction, can either decrease or increase the probability of contagion. The type of correlations observed in real networks actually act to reduce contagion.
\end{itemize}
The paper is organized as follows: In the next section we review the main features of the model
introduced in \cite{GK10}. We then consider the effect of heterogeneous degree distributions in section \ref{sfdegree}, and the effect of having a power law distribution for balance sheet size in section \ref{sectionsfassets}. Finally, we consider in section \ref{correlation} the effect of introducing correlations in the system and provide a summary of our findings in the last section.

\section{Review of the model}\label{model}
In this section we describe the model introduced in \cite{GK10} and review its main features.
The model describes a system of $N$ financial institutions (banks for brevity), each of them representing a node of a random network. In this context, links mimic interbank exposures and are therefore directed and weighted. Incoming links represent assets for a bank, while outgoing links are liabilities, so that the interbank liabilities of a given bank correspond to interbank assets for other banks.
Each bank $i$ is characterized by a balance sheet that comprises interbank assets $A_i^{IB}$, interbank liabilities $L_i^{IB}$, deposits $D_i$ and illiquid assets $A_i^{M}$. The condition for the bank to be solvent is therefore
\be
(1-\phi) A_i^{IB}+q A_i^{M}-L_i^{IB}-D_i>0,
\ee
where $\phi$ represent the fraction of banks to which $i$ is exposed that have defaulted and $q$ is the sale price of the illiquid assets.
In the following we will assume for simplicity that $q=1$.
In this case, the solvency condition can be expressed as
\be
\phi<\frac{K_i}{A_i^{IB}},
\ee
where $K_i=A_i^{IB}+A_i^{M}-L_i^{IB}-D_i$ is the bank's capital buffer.\\

\begin{center}
\begin{figure}[h]
\begin{center}
\includegraphics[width=10cm]{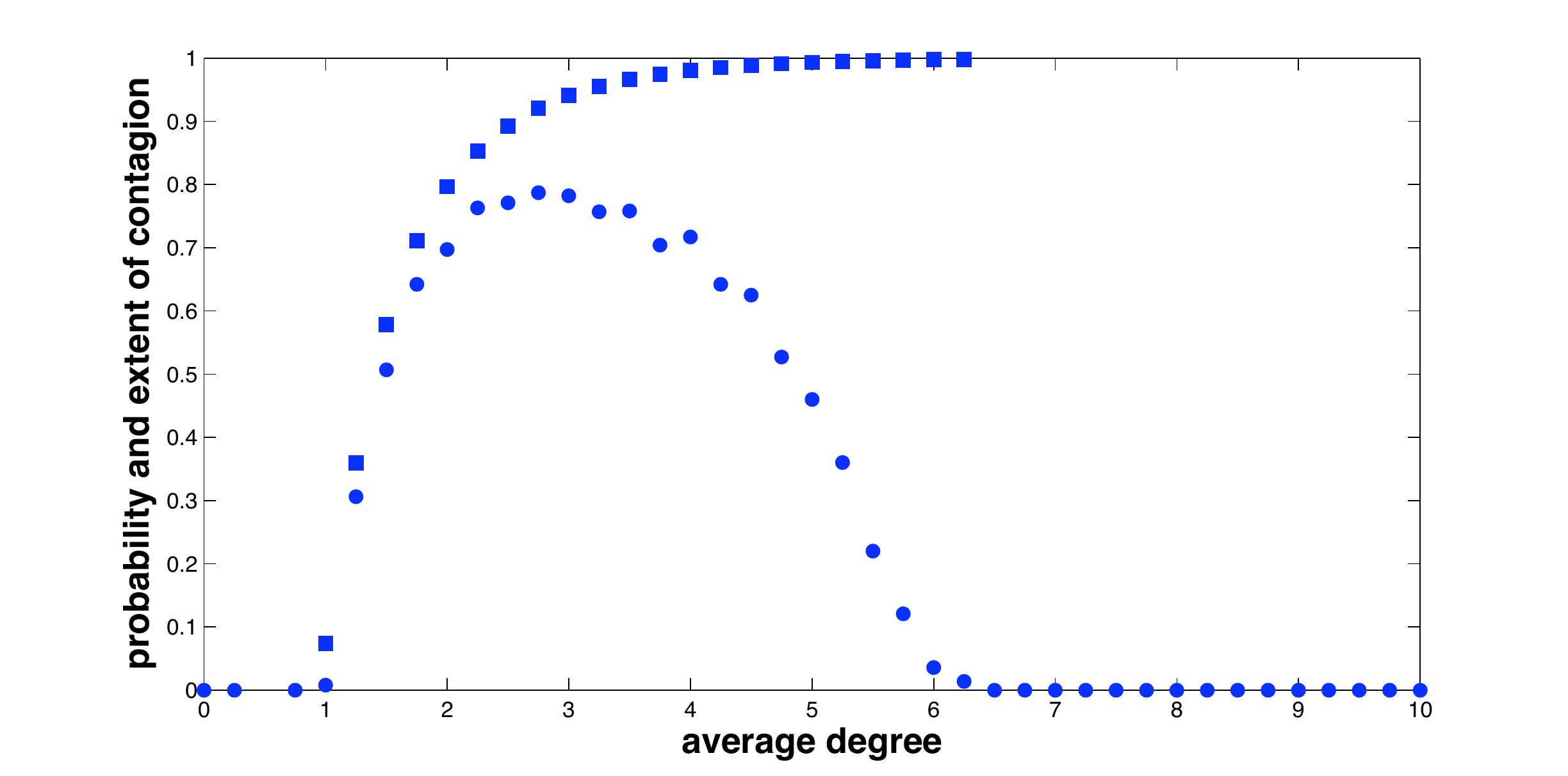}
\caption{\footnotesize {\textit{Contagion probability (circles) and conditional extent of contagion (squares) for Erd\"{o}s-Renyi random graphs as a function of the average degree. Results refer to  $1000$ simulations of networks with $10,000$ nodes. Banks balance sheet comprise $20\%$ of interbank assets, $4\%$ of capital buffer, $76\%$ of illiquid assets, and we set $q=1$. There is a window of connectivities with a finite probability of contagion. }}}\label{fig1}
\end{center}
\end{figure}
\end{center}

We are interested in understanding how the system responds to the bankruptcy of a single bank, and in particular when this results in contagion events in which a finite fraction of the banks are shut down.
In \cite{GK10}, generating functions techniques are exploited to give an analytical solution to the problem, and results from numerical simulations are presented as well.
The main results of \cite{GK10} that are going to be relevant for this paper are the following (see figure \ref{fig1}):
\begin{itemize}
\item As a function of the average connectivity, the system displays a window of connectivities for which the probability of contagion is finite. Links have a two-fold role in the system:  On one hand, they allow for contagion to spread across neighboring banks, on the other hand they imply that risks can be shared by many banks.
\item Increasing the capital buffer of banks reduces the contagion probability.
\item When the system is very connected (i.e. when the average degree is high), the system displays a robust yet fragile property.  The probability of contagion is very low, but in those instances where contagion happens the whole system is shut down. 
\end{itemize}
In \cite{GK10}, numerical results are presented for the case of Erd\"{o}s-Renyi random graphs and homogeneous distribution of assets.
In this paper, we intend to investigate how different topologies and different distributions of assets can affect the stability of the system. We also study the effect of different policies aimed at enhancing the stability of the system.

\begin{center}
\begin{figure}[h]
\begin{center}
\includegraphics[width=10cm]{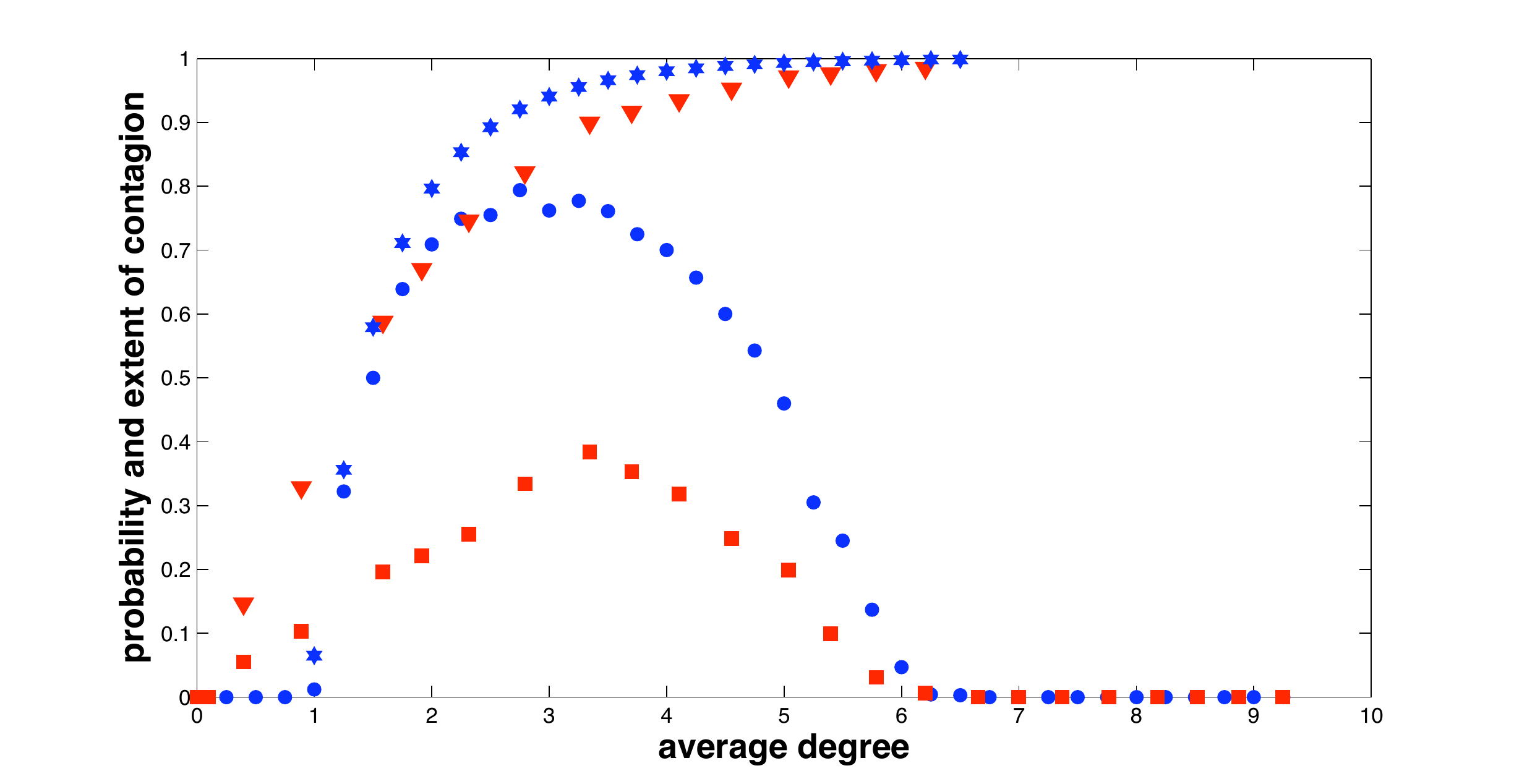}
\caption{\footnotesize {\textit{Blue circles: contagion probability for Erd\"{o}s-Renyi random netorks. Red squares: contagion probability  for scale free networks with $\gamma=3.0$. Blue stars: extent of contagion for Erd\"{o}s-Renyi random netorks. Red triangles: extent of contagion for scale free networks with $\gamma=3.0$.  Results refer to  $1000$ simulations of networks with $10000$ nodes. Scale free networks appear to be more resilient to contagion.} }}\label{fig2}
\end{center}
\end{figure}
\end{center}

\section{Heterogeneous degree distributions}\label{sfdegree}

Empirical evidence indicates that real financial networks exhibit heterogeneous degree distributions \cite{emp_italy,emp_italy2,emp_Austria}. The number of connections varies considerably across banks, with a few highly connected banks playing the role of hubs.  To understand how this affects financial stability we conducted numerical simulations of the above model using scale free networks, i.e. networks with a power law distribution of in and out degrees: $P(z)=z^{-\gamma}$. The distinctive feature of a scale free network is the existence of nodes with very different degree, and in particular the existence of hubs  with a large number of connections, 
property that can have a large impact on both static and dynamical properties of these systems\cite{DMG08,AB02,Bianconi02,CD09}.
In this section we will discuss how a scale free degree distribution affects the stability properties of the model introduced in section \ref{model}. In this paper, unless otherwise stated, we will consider  balance sheets comprising $76\%$  illiquid assets, $20\%$ interbank assets and a capital buffer $K_i=4\%$ for all the banks.
As in \cite{GK10} the contagion probability is computed as the fraction of runs for which more than $5\%$ of banks are shut down because of the initial failure of one single bank, and the extent of contagion is computed as the average fraction of bankruptcies conditional on the fact that at least $5\%$ of the system has been shut down. These quantities are the result of $1000$  runs, and
 are reported in figure \ref{fig2}, where we present a comparison between Erd\"{o}s-Renyi and scale free networks for the case where a random bank is initially shut down. 
The scale free topology enhances the stability of the system as far as the contagion probability is concerned, but doesn't significantly affect the extent of contagion and the robust yet fragile nature of the system.
\begin{center}
\begin{figure}[h]
\begin{center}
\includegraphics[width=10cm]{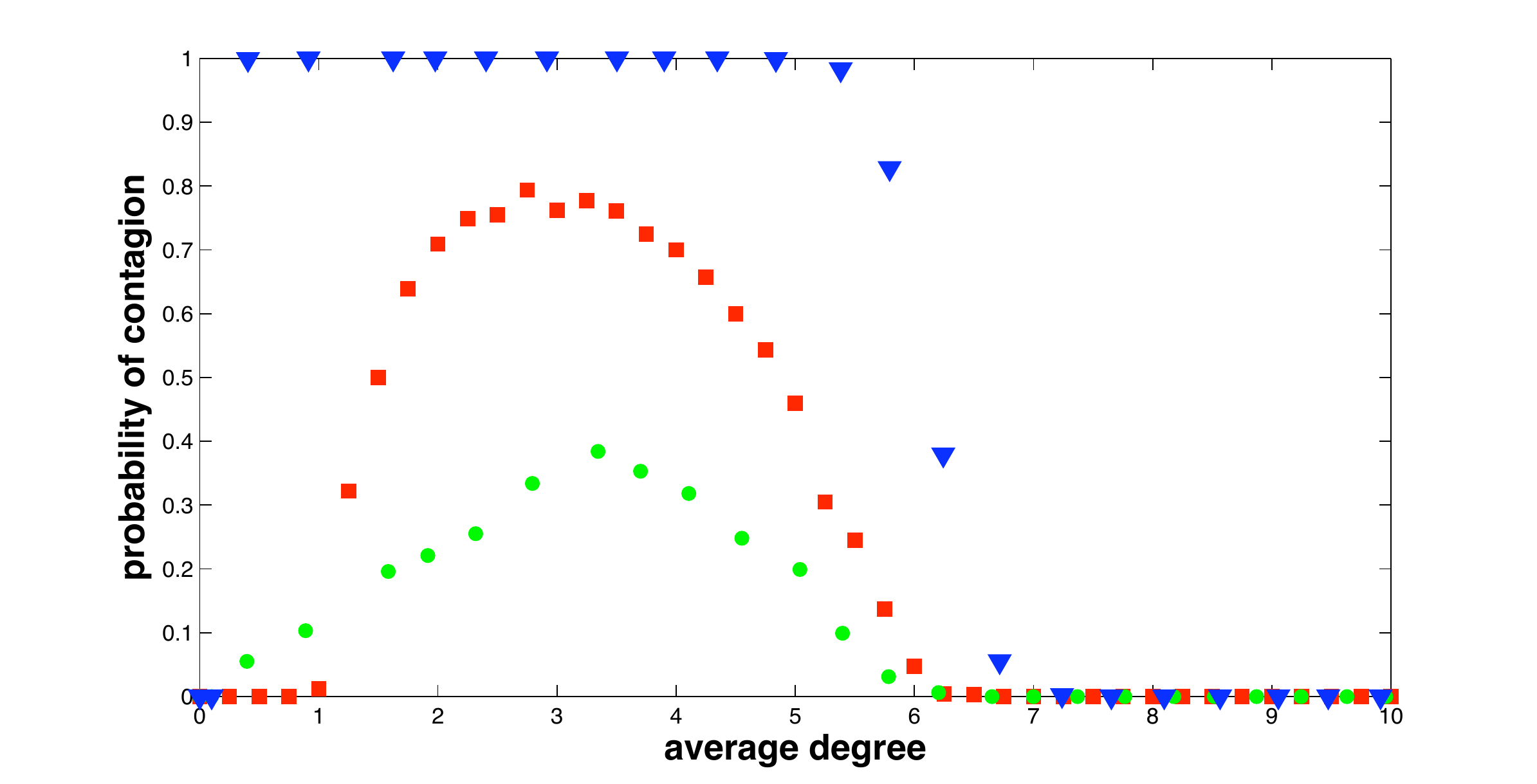}
\caption{\footnotesize {\textit{ Red squares: contagion probability for Erd\"{o}s-Renyi (red squares) and scale free (green circles) networks with $\gamma=3.0$ under random initial failure and for scale free networks  conditional on the initial failure of the highest connected node(blue triangles).  Results refer to  $1000$ simulations of networks with $10000$ nodes. On one hand scale free networks appear to be more resilient  than Erod\"{o}s-Renyi if the seed is chosen with uniform probability across nodes. On the other hand the probability of contagion in scale free networks is much higher if the seed is the most connected node.}}}\label{fig2b}
\end{center}
\end{figure}
\end{center}
\begin{center}
\begin{figure}[h]
\begin{center}
\includegraphics[width=10cm]{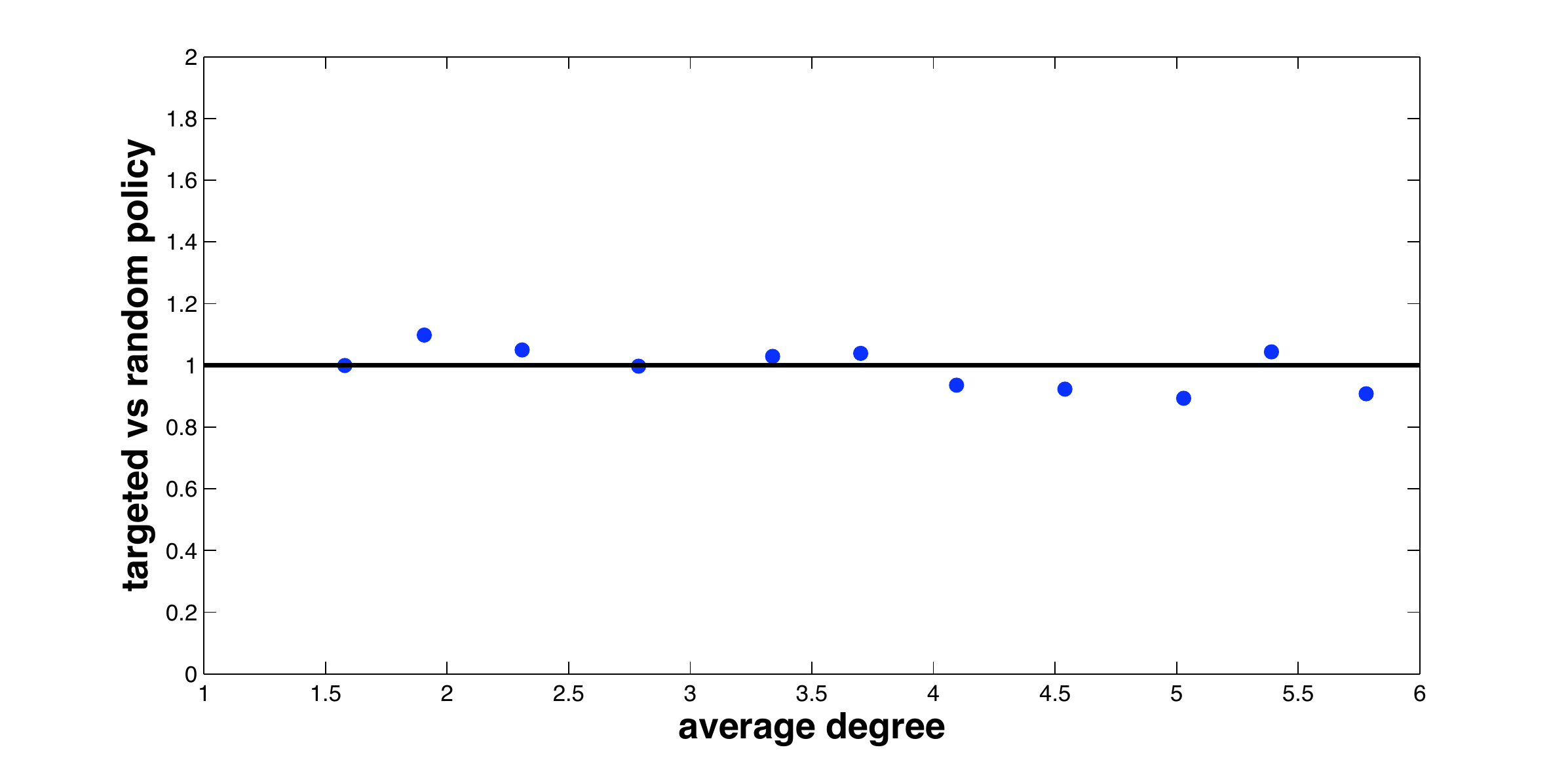}
\caption{\footnotesize {\textit{Ratio between contagion probability in presence of a targeted policy where the capital buffer of the $5\%$ most connected banks is increased and the contagion probability in the case where the capital buffer of $5\%$ random banks is increased for scale free random networks with $\gamma=3$.  Results refer to  $1000$ simulations of networks with $10000$ nodes. Such ratio oscillated around $1$ signaling  that, for the model we consider here, a targeted policy is not effective.}}}\label{fig2c}
\end{center}
\end{figure}
\end{center}
 \begin{center}
\begin{figure}[h]
\begin{center}
\includegraphics[width=10cm]{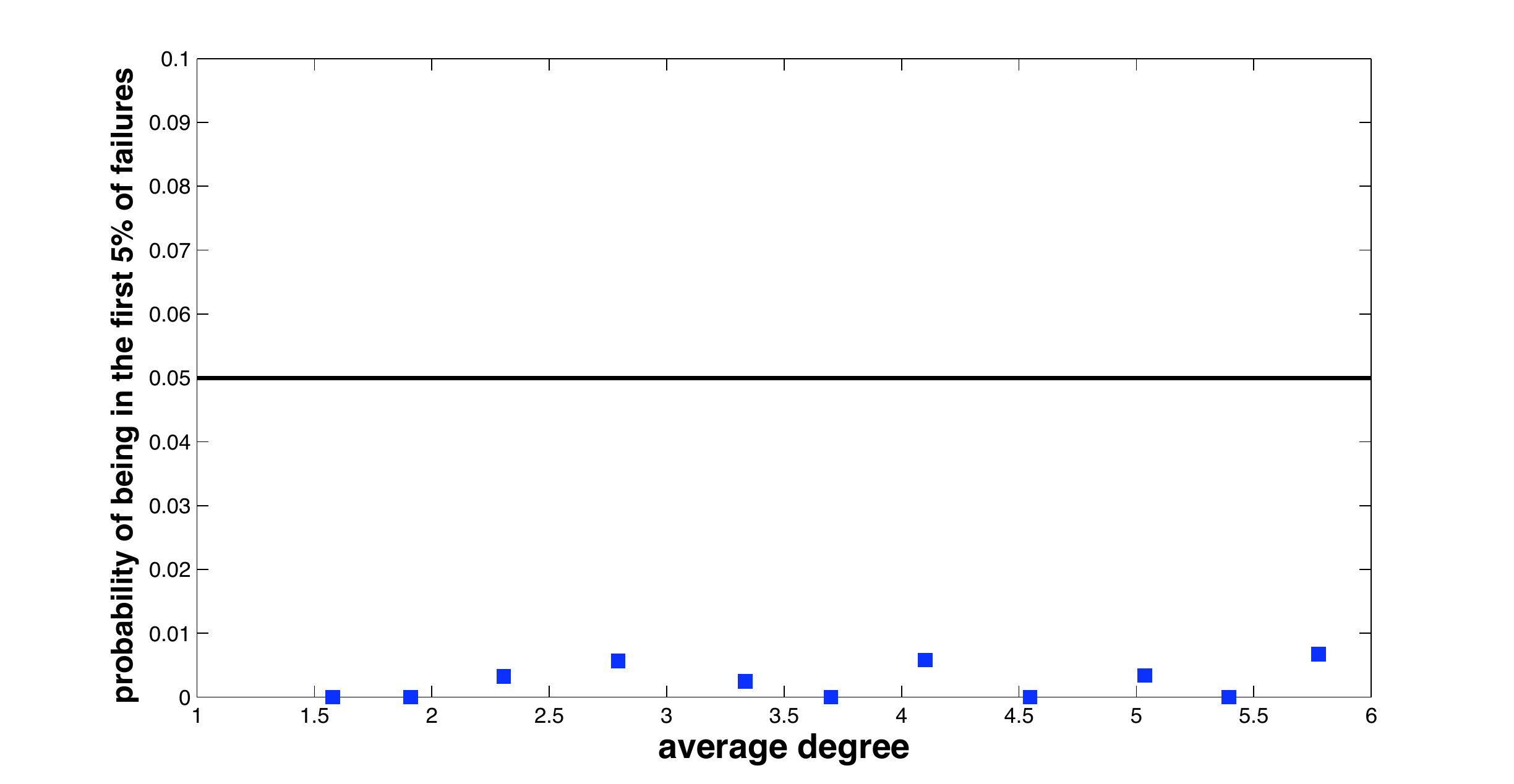}
\caption{\footnotesize {\textit{Probability that the most connected bank is within the first $5\%$ of failed banks for scale free networks with $\gamma=3.0$. The black line represents the reference level of $5\%$.  Results refer to  $1000$ simulations of networks with $1000$ nodes. The bankruptcy of the most connected bank always happens after a substantial number of other banks have already gone down.}}}\label{trigger}
\end{center}
\end{figure}
\end{center}
In the complex networks literature it has typically been seen that scale free networks are more robust than homogeneous ones with respect to random failures but more fragile in the case of a targeted attack \cite{AB02}.  To test whether this is true here as well, we look at the probability of contagion conditional on the initial failure of a high degree node.
We consider in particular the extreme case where the most connected node is selected to default. Results are reported in figure \ref{fig2b}, where we show that, as expected, scale free networks are more unstable with respect to targeted failures. 
This is in agreement with similar results obtained in \cite{CG11} for the average extent of contagion.  Again, as in the case of targeted attack there is no large effect  for the conditional extent of contagion.\\

The very fact that the probability of contagion for scale free networks is higher in the case of a targeted failure seems to suggests that banks with a large number of connections are of particular systemic importance. A natural question at this point is whether this fact can be used to introduce a policy of targeted reserve requirements that increase systemic stability. In other words, is it possible to decrease the probability of contagion by increasing the capital buffer of a few selected nodes?

Within the context of the model discussed here this doesn't seem to be the case. In figure \ref{fig2c} we compare the probability of contagion for a system where we increased  the capital buffer to $6\%$ of the $5\%$ of nodes with highest degree.  We compare this to a simulation where we increased the capital buffer in the same way for $5\%$ of banks chosen at random. As we see from the plot the ratio of the two probabilities oscillates around $1$, indicating that the targeted policy is not effective. 

This can be intuitively understood through the following argument: The weight of an interbank claim associated with an incoming link is for each bank inversely proportional to its in-degree, since we assumed interbank assets to be equally distributed among neighbors. This implies that the first banks to be affected by the initial random failures are going to be poorly connected banks, for which the importance of a single link is higher, while highly connected banks are going to be affected only at a later time. To confirm this intuition, we computed the probability that, given the failure of the most connected bank, such bankruptcy happens after at least $5\%$ of the other banks have been shut down. In a random setting, we would expect that the probability of going bankrupt within the first $5\%$ of banks is $5\%$. 
If, however, banks with high connectivity tend  on average to go bankrupt at a later stage, we expect the measured probability to be much lower than $5\%$, which is clearly the case for the most connected node (see figure \ref{trigger}).
Such behavior for threshold models has already been observed within the context of public opinion formation in \cite{WD07}, where it was shown that cascades of influence are not driven by a few very influential nodes, but by a critical mass of influenced individuals. 

Summarizing the results obtained in this section we can say that:
\begin{itemize}
\item Heterogeneous degree distributions decrease the probability of contagion triggered by a random failure.
\item Scale free networks are characterized by a few influential banks with a lot of connections, and the probability of contagion conditional on the failure of one of these influential nodes is higher than for homogeneous networks.
\item Within the context of the present model, a targeted policy where the capital buffer of a few ($<5\%$) banks is increased is not enough to further diminish the probability of contagion.
\item Although the probability of contagion is tamed by a heterogenous degree distribution, the conditional extent of contagion (i.e. the fraction of failed banks in those cases where at least $5\%$ of the system has gone down) is not significantly different with respect to the case of a homogeneous degree distribution. In particular, the robust yet fragile behavior highlighted in \cite{GK10} is preserved.
\end{itemize}

\section{Heterogeneous distribution of assets}\label{sectionsfassets}
In the previous section we considered networks made of interlinked balance sheets with heterogeneous connectivities.  Another important source of heterogeneity that has been empirically observed is that bank balance sheets are highly non-uniform, and appear to follow a power law distribution \cite{emp_Austria}. In this section we will focus on this property with the aim of understanding how this affects the stability of the system.
\begin{center}
\begin{figure}[h]
\begin{center}
\includegraphics[width=10cm]{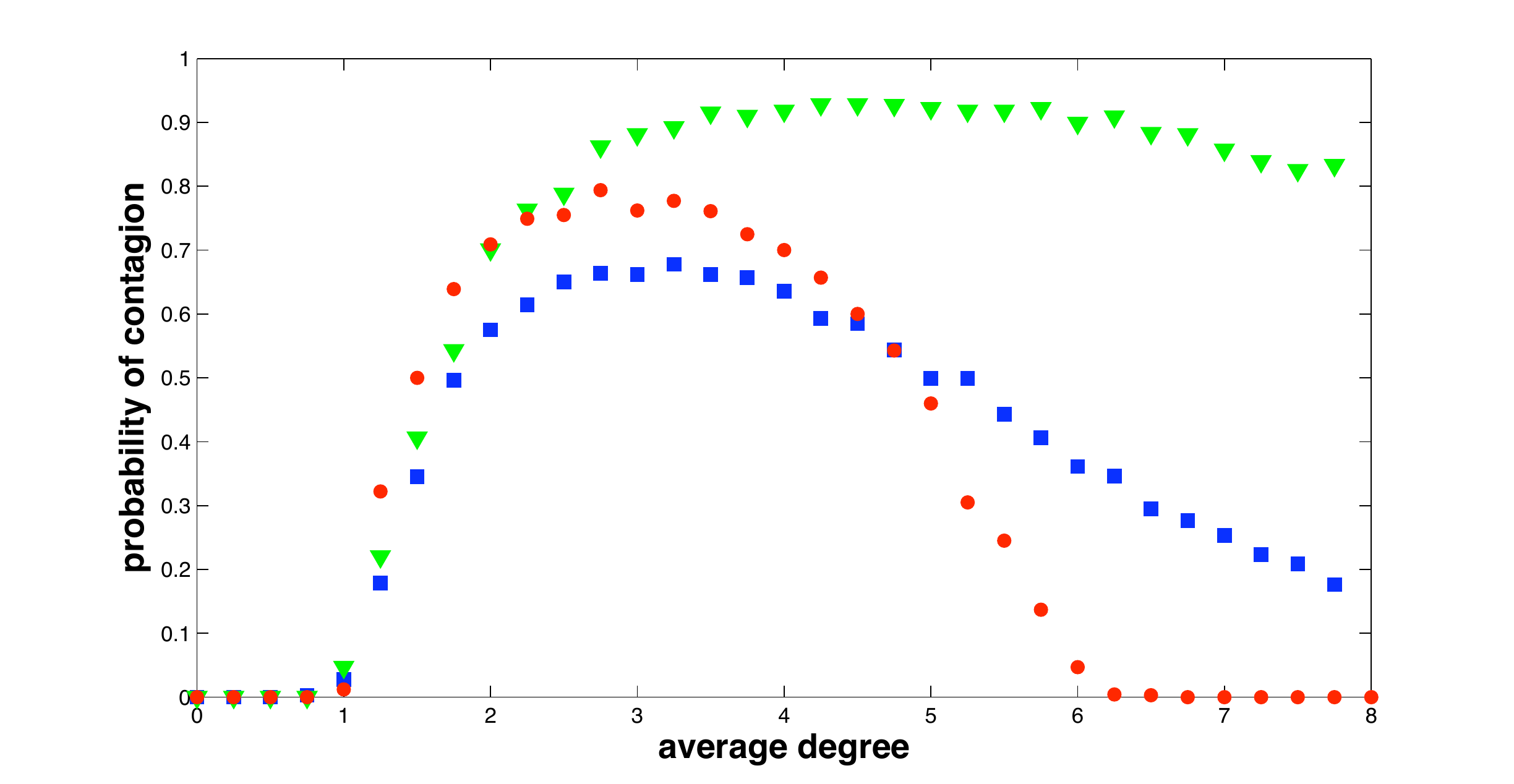}
\caption{\footnotesize {\textit{Red circles: contagion probability for Erd\"{o}s-Renyi  random graphs with uniform asset distribution. Blue squares: contagion probability for Erd\"{o}s-Renyi  random graphs with power law asset distribution under random initial failure.  Green triangles:   contagion probability for Erd\"{o}s-Renyi  random graphs with power law asset distribution conditional on the initial failure of the bank with the biggest balance sheet.   Results refer to  $1000$ simulations of networks with $1000$ nodes. The inefficient diversification induced by the heterogeneous asset distribution widens the contagion window. The situation is even worse when we consider the contagion probability conditional on the failure of the biggest bank.}}}\label{sfassets}
\end{center}
\end{figure}
\end{center}
\begin{center}
\begin{figure}[h]
\begin{center}
\includegraphics[width=10cm]{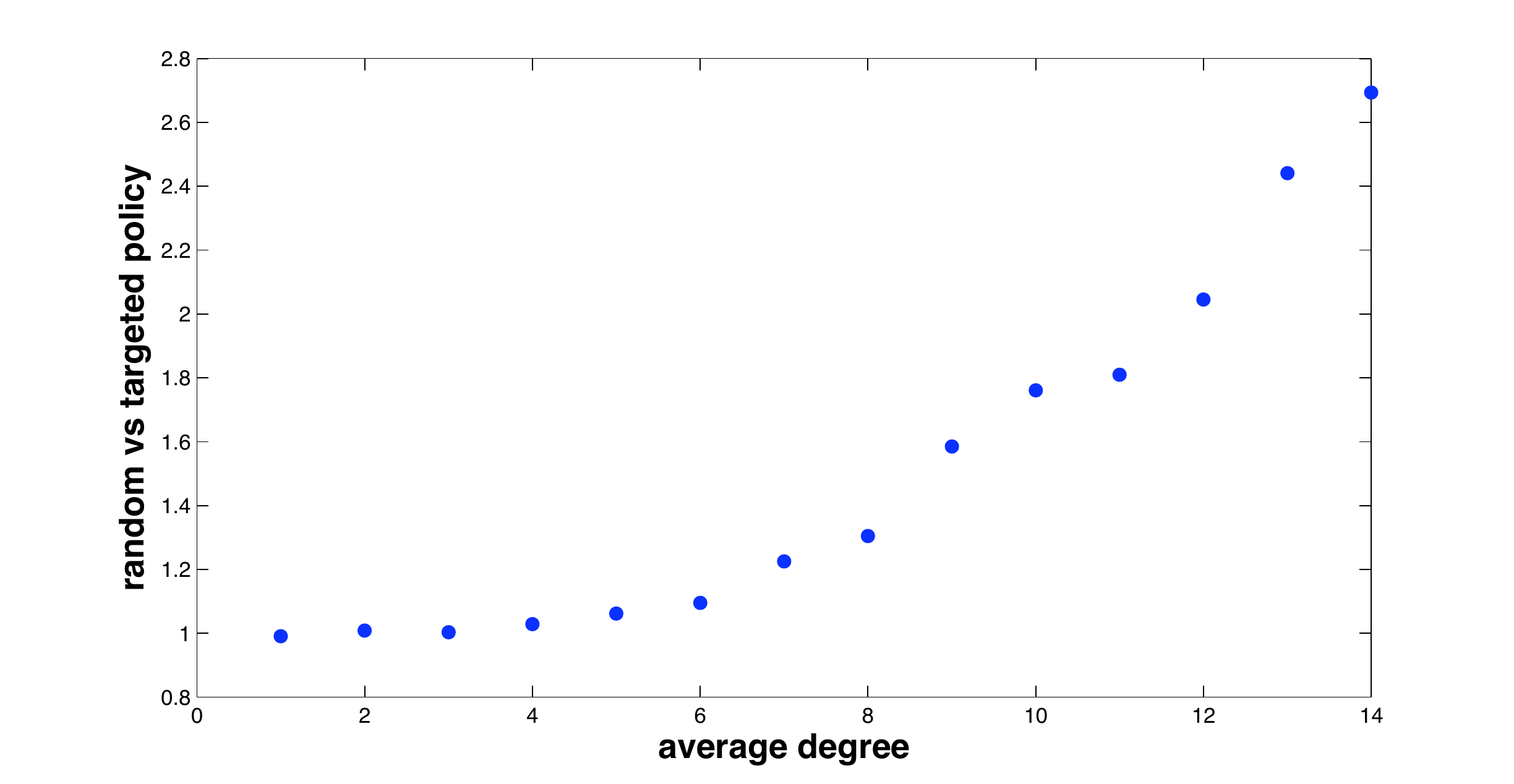}
\caption{\footnotesize {\textit{Ratio between the contagion probability when the capital buffer of random banks is increased vs. when the capital buffer of the $5\%$  biggest banks is increased, assuming Erd\"{o}s-Renyi  random networks.  Results are based on $10, 000$ different simulations of networks with $1000$ nodes each. The targeted policy is quite effective when average degree is large, where the probability of contagion can be reduced by more than $50\%$.}}}\label{sfassetspolicy}
\end{center}
\end{figure}
\end{center}
\begin{center}
\begin{figure}[h]
\begin{center}
\includegraphics[width=10cm]{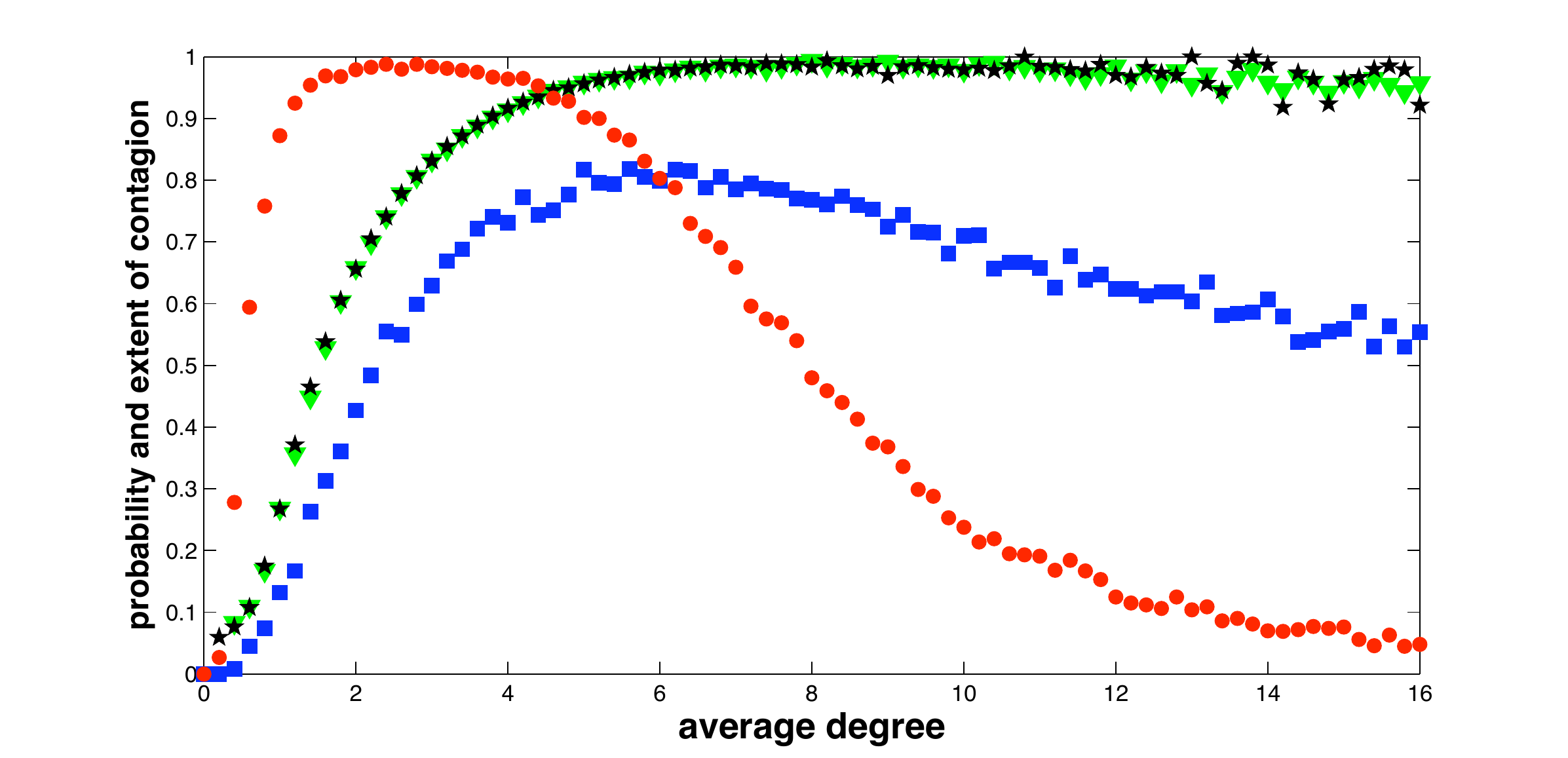}
\caption{\footnotesize {\textit{Red circles: contagion probability conditional on the initial failure of the most connected bank. Blue squares: contagion probability conditional on the initial failure of the bank with biggest balance sheet. Green triangles: conditional extent of contagion associated to the initial failure of the most connected bank. Black stars: conditional extent of contagion associated to the initial failure of the biggest bank.  Results refer to simulations of $10,000$ different networks with $1000$ nodes each. For low average connectivity the contagion probability is more dependent on connectivity than size.  For high average degree the opposite is true: the contagion probability conditional on the failure of the biggest bank is much higher than the contagion probability conditional on the failure of the most connected bank. No difference is seen for the corresponding extent of contagion. }}}\label{toobig}
\end{center}
\end{figure}
\end{center}
We consider a network of banks where the assets $A$ of a given bank are drawn from a distribution of the form $P(A)\propto A^{-\alpha}$. As before $20\%$ of these assets are in the form of interbank claims, but these assets are now distributed among in-neighbors according to the amount of assets held by neighboring banks. In this way we enforce, on average, a fixed ratio of the total amount of interbank assets and liabilities. The important difference with respect to the case of where all banks have the same assets is that the bank is not uniformly exposed to the failure of its neighbors. The result of this redistribution of exposures can be deleterious: a bank that would have been stable with respect to the failure of one neighbor in the case of uniform exposures can now go bankrupt if the wrong neighbor goes down. In other words a non uniform distribution of exposures implies less effective risk diversification, and we expect the probability of contagion to be affected.

This intuition is confirmed by the numerical simulations shown in figure \ref{sfassets}, where we show results obtained for the case $\alpha=2.5$ in Erd\"{o}s-Renyi random networks. We observe that the contagion window is wider in the case of a power law distribution of assets than it is for a uniform distribution of assets.  The inefficient allocation of exposures is highlighted if we consider the probability of contagion conditional on the fact that the biggest bank goes bankrupt (see figure \ref{sfassets}). As for the case of heterogenous degree distributions, we can ask whether the fact that a few big banks exist could be used to reduce the overall systemic risk through a policy of targeted reserve requirements. 
We then consider the situation where we increased to $6\%$ the capital buffer of $5\%$ banks with the highest  balance sheet size. The results are reported in figure \ref{sfassetspolicy}, where we compare this policy of targeted reserve requirement with a policy where we increased the capital buffer of $5\%$ of the banks chosen at random. We observe that, at least for high average degrees, a targeted policy can further reduce the contagion probability by more than $50\%$. This can actually be quite relevant if we think that this is the region where the robust yet fragile nature of the system is displayed, so that any contagion event can result in a failure of the whole system.\\

A topic of active discussion in recent years is whether institutions are of systemic importance because they are ``too big to fail'' or because they are ``to interconnected to fail'' \cite{tooint}.  We address this question here by
considering a network with power law distributions of both balance sheet size and degree and looking at what happens when the biggest vs the most connected bank fails. In order to distinguish the effects of size and connectivity we built the financial network without correlations between degree and size of banks (in other words the biggest bank is not necessary going to be the most connected). 

Results are reported in figure \ref{toobig}. We observe that the two properties dominate in different regimes.
For low average degree the probability of contagion due to the failure of the most connected bank is higher than that due to the failure of the biggest one.  For higher values of the degree, however, the opposite is true: size is actually more relevant than connectivity in terms of probability of contagion. Given that real networks appear to be in the region with high average connectivity, and that we showed that this is the regime where a policy of increasing the capital requirement of the biggest banks works, this would suggest that size is more important to reduce the probability of contagion. 

In summary we can say that:
\begin{itemize}
\item A power law distribution of balance sheet size induces a finite probability of contagion over a wider range of average connectivities than a uniform distribution.
\item A targeted policy where the capital requirement of biggest banks increases can be used to reduce the contagion probability in the regime of high average degree.
\item  The average degree of the network determines whether size or connectivity is more important.  When the degree is low the most connected banks are more important, but when the degree is high the size of a bank becomes more important.  Since real networks are in the high degree realm, this suggests that a "too big to fail policy" is more relevant.
\item The conditional extent of contagion is not significantly affected by heterogenous assets distributions.
\end{itemize}

\section{Correlated financial networks}\label{correlation}
In the previous sections we focused on the effect of heterogeneity on the stability of the system.  In this section we instead explore the effect of correlations between neighboring banks.  A property of real financial networks that has been empirically observed is the fact that banks with a few connections tend to form links with highly connected banks. This property, known as disassortative mixing, is actually common for technological networks, while the opposite is usually true for social networks \cite{AB02}. 
\begin{center}
\begin{figure}[h]
\begin{center}
\includegraphics[width=10cm]{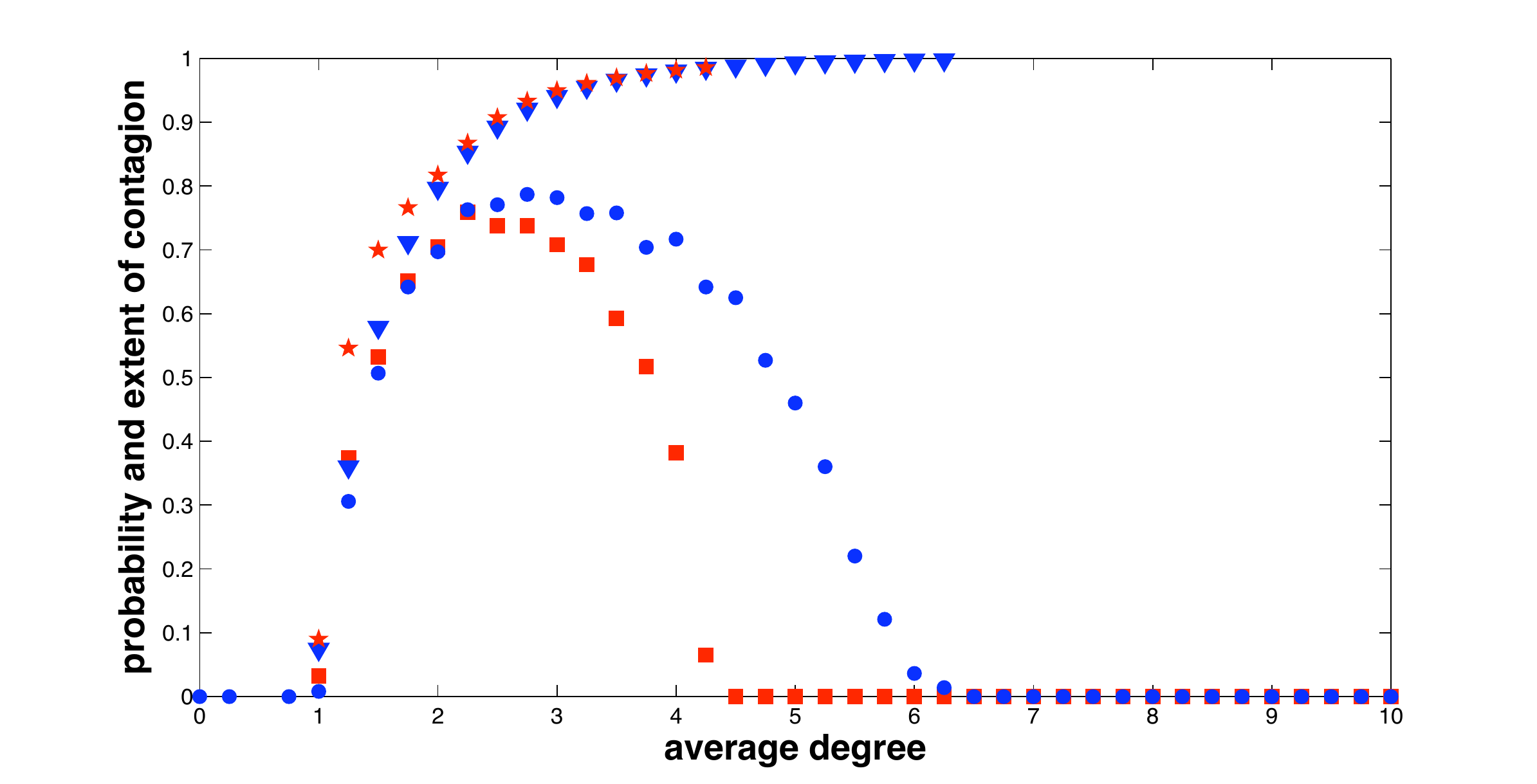}
\caption{\footnotesize {\textit{Blue circles: contagion probability for uncorrelated networks. Red squares: contagion probability for disassortative networks. Blue triangles:conditional extent of contagion for uncorrelated networks. Red stars: conditional extent of contagion for disassortative networks.  Results refer to  $1000$ simulations of networks with $10000$ nodes. The contagion window is smaller in the case of disassortative networks, but the conditional extent of contagion is always high in for high average degree.}}}\label{fig3}
\end{center}
\end{figure}
\end{center}

\begin{center}
\begin{figure}[h]
\begin{center}
\includegraphics[width=10cm]{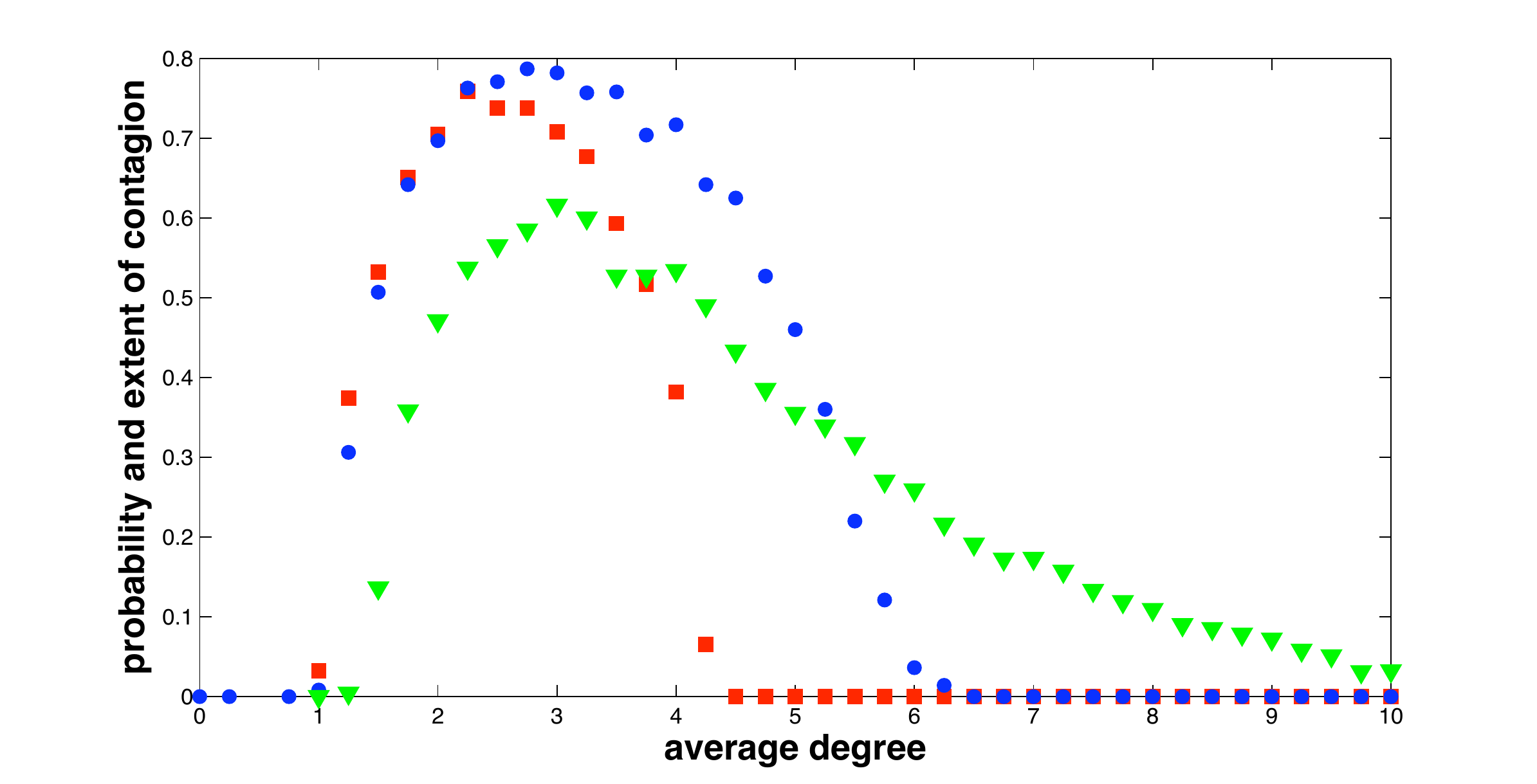}
\caption{\footnotesize {\textit{Blue circles: contagion probability for uncorrelated networks. Red squares: contagion probability for disassortative networks. Green triangles: contagion probability for assortative networks.  Results refer to simulations of $1000$ different networks each with $10,000$ nodes. Positive correlations among degrees widen the contagion window.}}}\label{fig3b}
\end{center}
\end{figure}
\end{center}

We want to understand the impact of disassortativity on the stability of the system. Since we would like to disentangle the effect of heterogeneity in the degree distribution from that of degree correlations, we consider here the case of homogeneous random graphs with correlations among degrees. Such networks have been generated starting with Erd\"{o}s-Renyi random graphs and then applying the rewiring method introduced in \cite{Noh07}. The method is essentially a Monte Carlo algorithm where the cost function to minimize  is 
\be
H(G)=\frac{J}{2}\sum_{i,j} a_{i,j}k_ik_j,
\ee
where $a_{i,j}=1$ if banks $i$ and $j$ are connected and $a_{i,j}=0$ otherwise, and $J$ is a constant that can be used to tune the level of assortativity:  With $J<0$ we will obtain a disassortative network, with $J>0$ an assortative network, and with $J=0$ an uncorrelated network. 

In figure \ref{fig3} we show the probability and extent of contagion for the case of disassortative ($J=-1$) and uncorrelated ($J = 0$) Erd\"{o}s-Renyi random graphs. We observe that such correlations tend to reduce the window where the probability of a systemic event is finite.
This reduced probability of contagion induced by disassortativity can be intuitively understood through the following reasoning:
In a disassortative network less connected nodes have a tendency to be connected with higher connected nodes. Since higher connected nodes are the most robust, with this network configuration higher connected nodes act as a screen reducing the failure probability of less connected nodes. The opposite is true for assortative networks. Indeed in this case weak nodes (i.e. poorly connected ones) tend to be connected among themselves, and the window of contagion is wider, as shown in figure \ref{fig3b}.

\section{Conclusions}
In this paper we have considered a model of financial contagion recently introduced in \cite{GK10}, and we have studied  how the stability of the system is affected by features empirically observed in real networks. In particular we have considered the effect of power law distributions in degree and balance sheet size, and correlations in degree.  The stability of the system has been tested according  to different failure protocols,namely, initial failure of a random bank, initial failure of the most connected bank, and initial failure of the bank with the biggest balance sheet. \\

In our numerical simulations we observe that a scale free network topology decreases the probability that a random failure can trigger a series of cascades affecting a finite fraction of the system. However, the contagion probability conditional on the initial failure of the most connected bank is significantly increased.
This is because a system with scale free topology is characterized by a few hubs and many less connected nodes. The failure of a hub potentially has an impact on many other nodes.
However, in the case of the failure of a random bank, the probability that the selected bank is a hub is quite small. The fact that a few banks have a high impact on the stability of system might suggest that a policy where the capital buffer of those banks is increased might enhance the stability of the system. In the framework of the model here considered, however, such a policy is ineffective in the case of cascades triggered by the initial failure of less connected banks. This is because highly connected banks have more means of risk diversification, so that they fail only after a larger number of other banks have already gone down.\\

Another realistic feature we incorporated in the model is a power law distribution for the size of bank balance sheets. Our numerical experiments reveal an increased window of contagion with respect to the case of homogeneous sizes.  This can be traced back to the fact that banks are not uniformly exposed to their counterparty.  The situation is clearly worse if we consider the probability of contagion triggered by the failure of the biggest bank, since the big banks represent the highest source of exposure for their creditors. At odds with the case of a scale free degree distribution, however, a targeted policy where the capital buffer of a few big banks is increased can significantly reduce the contagion probability, at least in the region of high average degree. This is particularly important since this is exactly the region where the system displays robust yet fragile behavior, meaning that if contagion happens the whole system will go down.\\

A comparison between the ``too big to fail'' and the ``too interconnected to fail'' point of views was done by comparing the probability of contagion conditional on the failure of the most connected vs the biggest bank for a system with power law degree distributions of connectivity vs. asset size.  The failure of highly connected banks has a stronger impact for networks with low average degree, while the failure of large banks has more effect for networks with high average degree. This, combined with the previous results on the effect of  targeted  policies, suggests that the ``too big too fail'' approach is more effective, at least in the framework of this model.\\

Finally we took a look at the effect of introducing correlations into financial network. We found that the type of degree correlations observed in real financial networks tend to reduce the probability of contagion. 

\section*{Acknowledgments}
This work was supported by the National Science Foundation under Grants
No. 0965673 and 1005075 and by the Sloan Foundation under the Grant ``Network Models of Systemic Risk". The authors would like to thank Pierpaolo Vivo for a careful reading of the manuscript.

\bibliography{ref}{}

\begin{thebibliography}{10}

\bibitem{GK10}
P.~Gai and S.~Kapadia, ``Contagion in financial networks,'' {\em \rm Proc. R.
  Soc. A}, vol.~466, no.~2120, pp.~2401--2423, 2010.

\bibitem{TCJ08}
M.~G. Crouhy, R.~A. Jarrow, and S.~M. Turnbull, ``The subprime credit crisis of
  2007,'' {\em \rm The Journal of Derivatives}, vol.~16, no.~1, pp.~81--110,
  2008.

\bibitem{emp_italy}
C.~Iazzetta and M.~Manna, ``The topology of the interbank market: developments
  in italy since 1990,'' {\em \rm Banca d'Italia Working paper}, no.~711, 2009.

\bibitem{emp_italy2}
G.~Iori, G.~D. Masi, O.~V. Precup, G.~Gabbi, and G.~Caldarelli, ``A network
  analysis of the italian overnight money market,'' {\em Journal of Economic
  Dynamics and Control}, vol.~32, no.~1, pp.~259--278, 2008.

\bibitem{emp_Austria}
M.~Boss, H.~Elsinger, M.~Summer, and S.~Thurner, ``The network topology of the
  interbank market,'' {\em \rm Quantitative Finance}, no.~4, p.~677, 2005.

\bibitem{emp_Brasil10}
E.~B. e~Santos and R.~Cont, ``The brazilian interbank network structure and
  systemic risk,'' Working Papers Series 219, Central Bank of Brazil, Research
  Department, Oct. 2010.

\bibitem{NYYA08}
E.~Nier, J.~Yang, T.~Yorulmazer, and A.~Alentorn, ``Network models and
  financial stability,'' {\em Journal of Economic Dynamics and Control},
  vol.~31, no.~6, pp.~2033--2060, 2007.

\bibitem{CM09}
C.~Cornand and G.~Moysan, ``Unstable financial networks,'' {\em \rm available
  at ssrn.com/abstract=1494233}, 2009.

\bibitem{HM11}
A.~G. Haldane and R.~M. May, ``Systemic risk in banking ecosystems,'' {\em \rm
  Nature}, no.~469, pp.~351--355, 2011.

\bibitem{ACM10}
H.~Amini, R.~Cont, and A.~Minca, ``Resilience to contagion in financial
  networks,'' {\em Available at SSRN: http://ssrn.com/abstract=1865997}, 2010.

\bibitem{AG01}
F.~Allen and D.~Gale, ``Financial contagion,'' {\em Journal of Political
  Economy}, vol.~108, pp.~1--33, February 2001.

\bibitem{Georg10}
C.-P. Georg, ``The effect of the interbank network structure on contagion and
  financial stability,'' Global Financial Markets Working Paper Series 12-2010,
  Friedrich-Schiller-University Jena, Oct. 2010.

\bibitem{Watts02}
D.~Watts, ``A simple model of global cascades on random networks,'' {\em \rm
  PNAS}, no.~99, pp.~5766--5771, 2002.

\bibitem{DMG08}
S.~N. Dorogovtsev, A.~V. Goltsev, and J.~F.~F. Mendes, ``Critical phenomena in
  complex networks,'' {\em Rev. Mod. Phys.}, vol.~80, pp.~1275--1335, Oct 2008.

\bibitem{AB02}
R.~Albert and A.-L. Barab\'asi, ``Statistical mechanics of complex networks,''
  {\em Rev. Mod. Phys.}, vol.~74, pp.~47--97, 2002.

\bibitem{Bianconi02}
G.~Bianconi, ``Mean field solution of the ising model on a
  {B}arab\'asi-{A}lbert network,'' {\em Phys. Lett. A}, vol.~303, pp.~166--168,
  2002.

\bibitem{CD09}
F.~Caccioli and L.~Dall'Asta, ``Non-equilibrium mean-field theories on
  scale-free networks,'' {\em J. Stat. Mech.}, p.~P10004, 2009.

\bibitem{CG11}
X.~Chen and A.~Ghate, ``Financial contagion on power law networks,'' {\em
  Available at SSRN: http://ssrn.com/abstract=1751143}, 2011.

\bibitem{WD07}
D.~J. Watts and P.~S. Dodds, ``Influentials, networks, and public opinion
  formation,'' {\em Journal of Consumer Research: An Interdisciplinary
  Quarterly}, vol.~34, no.~4, pp.~441--458, 2007.

\bibitem{tooint}
{Int. Monetary Fund}, ``Global financial stability report,'' tech. rep., 2009.

\bibitem{Noh07}
J.~D. Noh, ``Percolation transition in networks with degree-degree
  correlation,'' {\em Phys. Rev. E}, vol.~76, p.~026116, Aug 2007.

\end{thebibliography}
\bibliographystyle{ieeetr}

\end{document}